\documentclass{ws-p8-50x6-00}

\begin{document}

\title{ELECTROWEAK MATRIX ELEMENTS AT LARGE $N_c$:\break MATCHING QUARKS TO MESONS.}

\author{S. Peris}

\address{Fisica Teorica and IFAE, Univ. Autonoma de Barcelona, 08193 Barcelona,Spain\\
E-mail: peris@ifae.es}




\maketitle

\abstracts{I review some progress made on the problem of
calculating electroweak processes of mesons at low energy with
the use of an approximation to large-$N_c$ QCD which we call the
Minimal Hadronic Approximation. An update of results for the
matrix elements of the electroweak penguin operators $Q_7$ and
$Q_8$ is also given.}

\section{Introduction}\label{sec:introduction}
Due to the disparity in scales  $(M_K/M_W)^2 \sim 10^{-4}$ it
becomes very useful to employ this ratio as an expansion
parameter and construct an Effective Lagrangian in which the
heavy $W$ field is integrated out. In practice one treats as
heavy all particles above and including the charm quark. The
technique for making this construction is very well
known\cite{Buras}. According to this technique, after having
integrated out all the heavy degrees of freedom by shrinking the
corresponding propagators to a point (see Fig. 1), one is left
with a Lagrangian which is a linear combination of ten four-quark
operators involving only the light quark fields $u,d,s$ coupled
to the gluon and the photon. Denoting these operators by the
generic form $\mathcal{Q}_j(\mu) \sim \overline{q}\Gamma_j q \
\overline{q}\Gamma'_j q$, where $\Gamma_j$ and $\Gamma'_j$ are
some known\cite{Buras} Dirac matrices, one has
\begin{equation}\label{ope}
  \mathcal{L}_{eff}= i \overline{q} \not\! D q +
  \sum_{j=1}^{10}\ c_{j}(\mu) \ \mathcal{Q}_j(\mu)\ .
\end{equation}
Shrinking a massive propagator to a point modifies the
\emph{ultraviolet} properties of the theory and the Wilson
coefficients $c_j(\mu)$ fix this so that the physics is, however,
not changed. Therefore the Wilson coefficients only know about
short-distance physics and are given by a series expansion in
powers of $\alpha_s$ for scales $\mu$ in between $M_W$ and a
typical hadronic scale $\Lambda_{had}\sim \mathcal{O}(1\ GeV)$.
 Since this series has large coefficients of the form
 $\log M_W/ \Lambda_{had}$ one uses the
Renormalization Group to resum them. Presently, due to the effort
mainly of two groups\cite{Buras2,Martinelli}, we know the Wilson
coefficients up to the next-to-leading order.

\begin{figure}[t]
\begin{center}
\epsfxsize=30pc 
\epsfbox{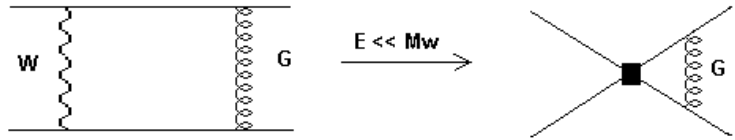} 
\caption{Schematic representation of the construction of an
Effective Lagrangian. $G$ and $W$  stand for a gluon and a $W$
exchange, respectively. \label{fig:one}}
\end{center}
\end{figure}

\subsection{The problem}\label{subsec:theproblem}

In fact, the separation made in the Lagrangian in Eq. (\ref{ope})
in terms of Wilson coefficients and operators must be defined by
some regularization  and depends on the conventions chosen like,
e.g., the scale $\mu$ used in the minimal subtraction, the value
of the  anticommutator $\{\gamma_{\mu},\gamma_{5}\}$ (i.e. whether
the NDR or HV prescription), the so-called evanescent
operators\cite{BurasWeisz}, etc... However, since the physical
result cannot depend on these conventions there has to be a
cancelation between the Wilson coefficients and the matrix
elements of $\mathcal{Q}_j$. This cancelation becomes a highly
nontrivial consistency check in any calculation.

Since kaons are light with respect to $\Lambda_{had}$ one still
writes a second Effective Lagrangian which is a  Chiral
Lagrangian, i.e. which is organized in powers of momenta and
masses of the light mesons according to chiral symmetry
\cite{Eduardo}. Meson matrix elements are then computed in terms
of masses and couplings of this Chiral Lagrangian such as, e.g.,
$f_\pi$.

A puzzle then arises. How can the convention dependence, e.g. of
$\gamma_{5}$, cancel between Wilson coefficients and matrix
elements of the Chiral Lagrangian which has no explicit
$\gamma_{5}$? Remarkably this problem has been plaguing all
analytic calculations of electroweak matrix elements up to date.

The key are the \emph{matching conditions}. These are conditions
on Green's functions which fix the value of the effective
couplings so as to keep the physics unaltered in the transition
from the fundamental to the Effective Lagrangian. Within a
perturbative context, this much was already understood. However,
perturbation theory cannot be valid if quarks are to make mesons.
The difficulty now lies in that we use two different languages to
describe the same physics: while at short distances we use quark
and gluons as our variables, at long distances we use meson
fields. Therefore the matching conditions entail a conceptually
new situation: one has to match an expansion in powers of
$\alpha_{s}$ coming from short distances to an expansion in
powers of meson momenta at long distances. In other words, in
order to make of the matching conditions a practical tool one has
to find a dictionary capable of translating a Green's function
from the language of quarks and gluons to the language of mesons.
In principle, if the solution to large-$N_c$ QCD were known, we
could use it as this dictionary. In practice, since this solution
is not known, we shall restrict ourselves to an approximation to
large-$N_c$ QCD that we shall call the ``Minimal Hadronic
Approximation''.

\begin{figure}[t]
\begin{center}
\epsfxsize=30pc 
\epsfbox{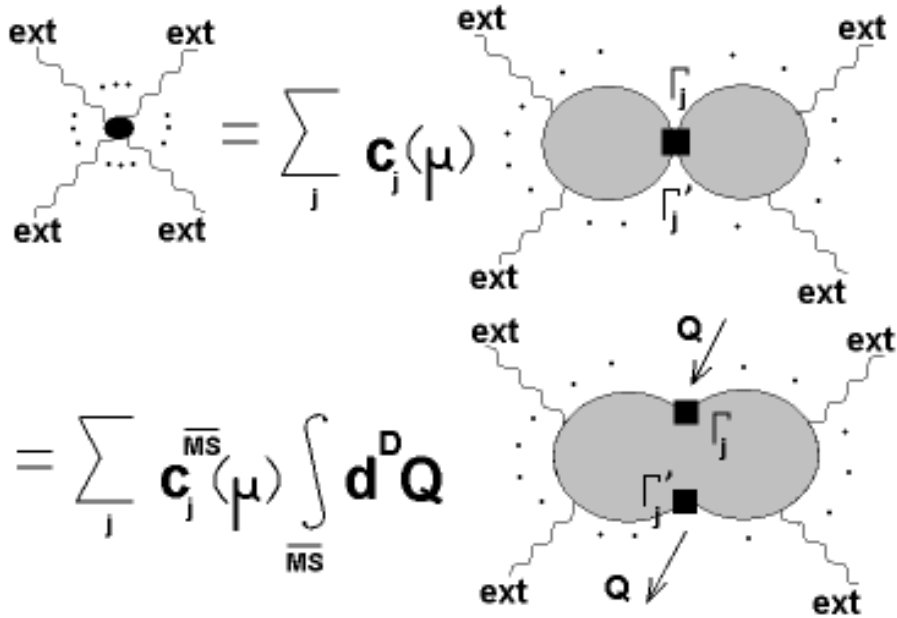} 
\caption{This figure represents a matching condition. Here ``ext''
stands for zero-momentum insertions of external fields. The gray
loops on the right-hand side signify the complete infinite set of
gluonic contributions which are leading at large $N_c$ (i.e.
planar diagrams with no internal quark loops). A possible
factorized contribution has been disregarded for simplicity. For
further explanations, see the text.\label{fig:two}}
\end{center}
\end{figure}

\subsection{Our solution: The Minimal Hadronic Approximation to large-$N_c$
QCD.}\label{subsec:mha}

What does the matching condition actually look like? The Chiral
Lagrangian is generically given by a function
$\mathcal{L}_{chiral}(D_{\mu}U)$, where $D_{\mu}U$ represents the
covariant derivative in the presence of external fields acting on
the usual exponential of the Goldstone fields. (External fields
are useful for identifying Green's functions.) On the other hand,
the quark-gluon Effective Lagrangian is given by Eq. (\ref{ope})
where the covariant derivative $D_{\mu}$ also contains the same
external fields as $\mathcal{L}_{chiral}$ plus the gluon, which is
certainly dynamic. In this situation the matching conditions can
be expressed pictorially as in Fig. 2.

This Figure expresses the fact that, at large $N_c$, coupling
constants in the Chiral Lagrangian contribute to a given Green's
function at tree level (left-hand side of Fig. 2). The higher the
number of external fields, the higher the order in the chiral
expansion. The matching condition demands that this Green's
function equals that obtained when the Lagrangian of Eq.
(\ref{ope}) is used (i.e. first graph on the right-hand side of
Fig. 2). This involves the insertion in the Green's function of
the \emph{local} four-quark operators in the Lagrangian
(\ref{ope}) which is actually a mathematically ill-defined
operation\footnote{A possible factorized contribution has been
disregarded here for simplicity.}. To appreciate the need to
define this \emph{locality} one can reexpress this contribution by
integrating over a \emph{euclidean} momentum $Q$ in the loop; and
this is divergent. Consequently a prescription has to be given,
and here is where the conventions used for the Wilson coefficients
come in. I have tried to represent this crucial point in the
second term on the right-hand side of Fig. 2 by the script
``$\overline{\mathrm{MS}}$'' on both the coefficients
$c_j^{\overline{\mathrm{MS}}}(\mu)$ and the $D-$dimensional
integral $\int_{\overline{\mathrm{MS}}} d^{D}Q$. The lesson from
Fig. 2 is that at leading order in $1/N_c$ the couplings of the
Chiral Lagrangian are given by $D-$dimensional integrals over an
euclidean momentum $Q$ of QCD Green's functions at large $N_c$,
with zero-momentum insertions of external fields and in the
\emph{forward} limit for $Q$. Let's call these QCD Green's
functions generically $\mathcal{G}(Q^2)$. If $\mathcal{G}(Q^2)$
was known for the full range of momentum $Q$ one could go ahead
and do the integral. The actual situation, of course, is that
hardly ever does one have all this information. Nevertheless, one
does have at least the chiral expansion of $\mathcal{G}(Q^2)$ for
low values of $Q^2$ and also the large-$Q^2$ expansion of
$\mathcal{G}(Q^2)$ given by its OPE, so that the problem becomes
rather
 how to build an interpolating function between these two regimes.
This interpolating function is what we have termed the ``Minimal
Hadronic Approximation'' (MHA) to Large-Nc QCD. It turns out that
some versions of Vector Meson Dominance are particular examples of
MHA, but they not always coincide.

To be specific, let's imagine that we want to construct the MHA
to the Green's function $\Pi_{LR}(Q^2)$ in the chiral and
large-$N_c$ limits. This function is defined as
$\Pi^{\mu\nu}_{LR}(q) = (q^{\mu} q^{\nu} - q^2 g^{\mu\nu})
\Pi_{LR}(q^2\equiv -Q^2)$, where
\begin{equation}\label{PiLR}
  \Pi^{\mu\nu}_{LR}(q)= 2 i\int d^{4}x\ e^{iqx}
  \langle 0|T\{L^{\mu}(x) R^{\nu}(0)^{\dag}\} |0\rangle \ ,
\end{equation}
and $R^{\mu}(L^{\mu})=\overline{d}(x)\gamma^{\mu}
\frac{(1\pm\gamma_{5})}{2} u(x)$. About $Q^2=0$ chiral
perturbation theory yields a Laurent series in $Q^2$,
\begin{equation}\label{chpt}
  \Pi_{LR}(Q^2)\approx - \frac{f_{\pi}^2}{Q^2}- 4 L_{10}+
  \mathcal{O}(Q^2)\ ,
\end{equation}
while about $Q^2=\infty$ the OPE is given by a (possibly
asymptotic) series in inverse powers of $Q^2$,
\begin{equation}\label{opeexp}
  \Pi_{LR}(Q^2)\approx \frac{h_1}{Q^2}+ \frac{h_2}{Q^4}+
  \frac{h_3}{Q^6}+ \mathcal{O}(\frac{1}{Q^8})\ ,
\end{equation}
with $h_1=h_2=0$ and $h_3=4 \pi \alpha_{s} \langle\overline{\psi}
\psi \rangle^2(1 + \mathcal{O}(\alpha_s \log Q^2))$. In order to
achieve matching with the Wilson coefficients at the
next-to-leading order we do not need to consider the
$\mathcal{O}(\alpha_s \log Q^2)$ in $h_3$\footnote{I thank A. Pich
for discussions on this point.}.

On the other hand, general properties of large-$N_c$
QCD\cite{thooft} tell us that $\Pi_{LR}(Q^2)$ is a meromorphic
function given by
\begin{equation}\label{largeNc}
  \Pi_{LR}(Q^2)=\sum_{V}^{\infty}\frac{f_V M_V^2}{Q^2+M_V^2} -
    \sum_{A}^{\infty}\frac{f_A
    M_A^2}{Q^2+M_A^2}-\frac{f_{\pi}^2}{Q^2}\ .
\end{equation}
Obviously, dealing with infinite sums is not a simple matter,
particularly when the poles and the residues are unknown. The
Minimal Hadronic Approximation is defined by keeping only a finite
number of resonances in these sums, whose residues and masses are
fixed by  matching to the first few terms of both the chiral  and
the OPE expansions, Eqs. (\ref{chpt},\ref{opeexp}). Once this is
done we have what is known in Mathematics as a rational
approximant. This is an interpolating function which is the ratio
of two polynomials and that, by construction, has the same low-
and high-$Q^2$ behavior as the full $\Pi_{LR}(Q^2)$.
 There is no
a priori condition on how many terms in both the chiral and the
OPE expansions one has to match. A sensible choice should
probably be made considering whether the particular observable
one is looking at weights more the low or the high-$Q^2$ tail; in
practice, however, one is limited by the availability of the
terms in these two expansions. Modulo this practical limitation,
the approximation is clearly systematic and well defined.

Therefore, keeping finite sums in Eq. (\ref{largeNc}), the
matching conditions  for our MHA read

\bea
    -4 L_{10} =  \sum_{V}^{N_V} f_V^2 &-&\sum_A^{N_A} f_A^2 \label{matchone}\\
    0=f_{\pi}^2+ \sum_A^{N_A} f_A^2 M_A^2  -  \sum_V^{N_V} f_V^2
    M_V^2 \quad &,& \quad
    0=\sum_A^{N_A} f_A^2 M_A^4 - \sum_V^{N_V} f_V^2 M_V^4 \label{matchthree} \\
    4 \pi \alpha_{s}\langle\overline{\psi}\psi\rangle^2 =
    \sum_A^{N_A} f_A^2 M_A^6& - &\sum_V^{N_V} f_V^2 M_V^6
    \label{matchfour}
    \ ,
    \eea
where as many terms from  Eqs. (\ref{chpt},\ref{opeexp}) as needed
are understood in order to fix the resonance parameters.
  Eq. (\ref{matchone}) is the statement of resonance
dominance considered by Ecker et al.\cite{Ecker}. Eqs.
(\ref{matchthree}) are nothing but a generalization of the
celebrated Weinberg sum rules\cite{Weinberg} and Eq.
(\ref{matchfour}) was first considered by Knecht and de
Rafael\cite{Knecht} and is the first page of the dictionary we
have been looking for: it relates an expression written in terms
of quarks and $\alpha_s$ to an expression in terms of
mesons\footnote{In actual fact what appears in Eq.
(\ref{matchfour}) is the four-quark condensate. Here we have
taken the strict large-$N_c$ limit and factorized. This may not
always be very precise numerically; see the last section for a
potential example of this.}. In the next section we shall see its
usefulness. Finally, our $\Pi_{LR}(Q^2)$ in the MHA is like that
in Eq. (\ref{largeNc}) but with a finite set of resonances in the
sums whose residues and poles satisfy Eqs.
(\ref{matchone},\ref{matchfour}).

How well does the MHA work ?  In the case of $\Pi_{LR}(Q^2)$ even
only one vector and one axial (plus the pion) does quite well in
a comparison with ALEPH data\cite{Boris}.

\section{An instructive example: the $\pi^+-\pi^0$ electroweak mass
difference.}\label{sec:applications}

Pions are massless in the chiral limit provided the electroweak
interactions are switched off. However, the effective operator
\begin{equation}\label{mass}
  \mathcal{L}_{Chiral}= e^2\ C\ \mathrm{Tr}\left(Q_R U Q_L
  U^{\dag}\right)= -\ \frac{2 e^2 C}{f_{\pi}^2} \pi^+ \pi^- + \cdots\ ,
\end{equation}
where $Q_L=Q_R=\mathrm{diag}(2/3,-1/3,-1/3)$ and $e$ the electric
charge, shows that charged pions do pick up a mass for $e\neq 0$.

It is useful to think of the matrices $Q_{L,R}$ as arbitrary
external fields. If you expand the exponential $U$ in the operator
(\ref{mass}) and keep only the unity, this operator gives rise to
a ``mixed mass term'' between the field $Q_L$ and the field
$Q_R$. Therefore $C$ can be regarded as a coupling between the
two external fields $Q_L$ and $Q_R$ in this Chiral Lagrangian,
just like we discussed in the general introduction. Can one
determine $C$? The answer is yes. Following our previous
discussion of Fig. 2, it is given by a matching condition which
in this case reads\cite{pi}
\begin{equation}\label{C}
  C=\frac{3}{32\pi^2} \int_{0}^{\infty} dQ^2\ \left\{1-
  \frac{Q^2}{Q^2+M_Z^2}\right\}\ \left(-Q^2 \Pi_{LR}(Q^2)\right)\
  ,
\end{equation}
where the unity in $\{...\}$ above comes from the photon and the
$\frac{Q^2}{Q^2+M_Z^2}$ from the $Z$ propagator. Therefore $C$ is
of $\mathcal{O}(N_c)$. The $\Pi_{LR}(Q^2)$ function stems from the
fact that the external fields $Q_{L,R}$ couple to the left- and
right-handed quark currents (see Eq. (\ref{Zeff}) below). Since
charged currents are fully lefthanded, there can be no $W$
contribution to the $Q_L \times Q_R$ operator in Eq.
(\ref{mass}). In Eq. (\ref{C}), unlike Fig. 2, there are no
Wilson coefficients because the $Z$ propagator has not been
shrunk to a point yet, although it will be soon
\footnote{Obviously this cannot be done with the photon
propagator.}. Therefore $C$ is known if $\Pi_{LR}(Q^2)$ is known
over the full range of momentum, as we discussed in the
introduction. Now we may approximate this function by its MHA
defined as Eq. (\ref{largeNc}) with only one $V$ and one $A$
instead of the infinite sums. The net result is that in this MHA
one finds the amazingly simple expression
\begin{equation}\label{PiMHA}
  -Q^2 \Pi_{LR}(Q^2) =\frac{f_{\pi}^2 M_V^2 M_A^2}{(Q^2+M_V^2)
  (Q^2+M_A^2)}\ ,
\end{equation}
with which the coupling constant $C$ in Eq. (\ref{mass})can be
computed explicitly as \bea\label{CMHA}
  C_{\mathrm{MHA}}=\frac{3}{32 \pi^2} f_{\pi}^2 M_V^2&&\Bigg[\frac{M_A^2}{M_A^2-M_V^2}
  \log\frac{M_A^2}{M_V^2} \qquad\nonumber \\
  &&\quad - \frac{M_A^2}{M_Z^2}\left(\log\frac{M_Z^2}{M_V^2}-
  \frac{M_A^2}{M_A^2-M_V^2}\log\frac{M_A^2}{M_V^2}\right)\Bigg]\ .
  \eea
Assuming ${f_{\pi}}^{\chi-\mathrm{limit}}=87\pm 3.5\ MeV$ and
$M_V=748\pm 29\ MeV$ are large-$N_c$ values\footnote{These values
are extracted from a comparison of Aleph data with this very same
MHA\cite{Boris}. }, Eqs (\ref{matchone}-\ref{matchfour}) are
overconstrained. Since the four-quark condensate (see footnote on
page 6) is not very well known, in practice we use only Eqs.
(\ref{matchone}-\ref{matchthree}) as true constraints and we take
Eq. (\ref{matchfour}) as a prediction for the condensate. Notice
that these equations give $M_A$ in terms of $f_{\pi},M_V$ and
$L_{10}$. It turns out that this leads to $M_V^2/M_A^2 \simeq
0.50\pm 0.06$ (corresponding to the experimental number for
$L_{10}\simeq L_{10}(\mu=M_{\rho})=-(5.1\pm 0.2)\times 10^-3$),
which is how the celebrated Weinberg result\cite{Weinberg}
appears in our approach. Numerically this shifts the charged pion
mass by $\sim 4.88\ MeV$ to be compared with $4.5936(5) \ MeV$
which is the experimental number. The Z correction amounts only
to a relative $\sim 0.1 \% $ of the total contribution. One
expects a rough $30 \%$ to be a fair estimate of the error in the
result of Eq. (\ref{CMHA}). The photon contribution coincides
with the classic result by Low et al.\cite{Low}.

Our interest in the Z contribution was not really its numerical
result but the following conceptual lesson that can be drawn from
it\cite{pi}. Let's imagine that, as is usually done, one first
integrates the Z out by shrinking its propagator to a point. In
this case, the Effective Lagrangian due to Z exchange is given by
the four-quark operator
\begin{equation}\label{Zeff}
  \mathcal{L}_{eff}=\frac{e^2}{M_Z^2}\
  \left\{ Q_{LR}\equiv\ \overline{q}_L\gamma_{\mu}
  Q_L q_L\ \overline{q}_R\gamma^{\mu} Q_R q_R\ \right\}+ \cdots
\end{equation}
where the dots stand for some other terms which are irrelevant in
what follows. According to the Effective Lagrangian technique,
this operator is valid at the $M_Z$ scale. As one runs it down to
a  scale $\mu\sim \Lambda_{had}$ the four-quark operator
$Q_{LR}(\mu)$ mixes into the operator
\begin{equation}\label{D}
D_{LR}(\mu)\equiv (Q_L)_{ij}\
 (Q_R)_{kl}\ \overline{q'^i_L}  q^k_R(x)\ \overline{q^l_R} q'^j_L(x) \ ,
 \end{equation}
according to (I keep only the one-loop leading log for
simplicity):
\begin{equation}\label{mixing}
  Q_{LR}(M_Z^2)=Q_{LR}(\mu)- \frac{3}{2} \frac{\alpha_s}{\pi}
  \log\frac{M_Z^2}{\mu^2} D_{LR}(\mu^2)\ .
\end{equation}
The $\pi^+$ mass can be defined through the matrix element of the
Lagrangian (\ref{mass}),  $<\pi^+|\mathcal{L}_{Chiral}|\pi^+>$,
and because of Eqs. (\ref{Zeff},\ref{mixing}), it receives now two
contributions. Firstly let us consider  the one that comes from
$Q_{LR}(\mu)$. Closing the quarks in a single loop produces a
$Q_L\times Q_R$ term whose coefficient is again the function
$\Pi_{LR}$. This contribution has a similar structure as the
electromagnetic interaction discussed before, i.e.
\begin{equation}\label{firstpart}
  C(Q_{LR})=- \frac{3}{32 \pi^2}\int_{0}^{\infty}dQ^2\
  \frac{Q^2}{M_Z^2}\ \left(-Q^2 \Pi_{LR}(Q^2)\right)
\end{equation}
but, because it lacks the photon propagator, it is actually
divergent and one has to regularize the integral. It coincides
with the $Z$ contribution in Eq. (\ref{C}) if one takes bluntly
the limit $M_Z\rightarrow \infty$ inside the integral. Going back
to a general MHA with $N_{V,A}$ resonances it reads
\begin{equation}\label{firstpartMHA}
  C(Q_{LR}(\mu))=\frac{3}{32 \pi^2}
  \left(\sum_{A}^{N_A} f_A^2 M_A^6 \log\frac{M_A^2}{\mu^2} -
    \sum_{V}^{N_V} f_V^2 M_V^6 \log\frac{M_V^2}{\mu^2}\right) \ .
\end{equation}
The contribution to $C$ from the operator $D_{LR}(\mu)$ is
simpler due to the explicit power of $\alpha_s$ appearing in
(\ref{mixing}), which allows one to use factorization at
$\mathcal{O}(N_c)$ to extract the $Q_L\times Q_R$ contribution.
It is simply $(1/4) <\overline{\psi}\psi>^2$. Together with the
Wilson coefficient in Eq. (\ref{mixing}) this yields
\begin{equation}\label{secondpart}
  C(D_{LR}(\mu))=- \frac{3}{2} \frac{\alpha_s}{\pi}
  \log\frac{M_Z^2}{\mu^2}\ \frac{1}{4} <\overline{\psi}\psi>^2\ .
\end{equation}
The total result is of course $C=C(Q_{LR}(\mu))+C(D_{LR}(\mu))$
and one observes here the usual difficulty I discussed in the
introduction: the necessary cancelation of $\mu$ in $C$ is not
seen. The problem is deeply related to the fact that one piece
depends on $\alpha_s$ and comes from some perturbative running
 Eq. (\ref{mixing}) whereas the other comes from some matrix
element computed with mesons (Eq. \ref{firstpartMHA}). However
when the matching condition (\ref{matchfour}) is recalled all
pieces fall into place and one can explicitly see the expected
cancelation of $\mu$ in the final result for $C$, which is again
given by the full $Z$ propagator in Eq. (\ref{C}). As promised,
the
 condition (\ref{matchfour}) has played the role of a dictionary\footnote{See
 our calculation of $B_K$ for a more
sophisticated application of MHA showing explicit scheme
independence at the next-to-leading-log
 level\cite{BK}.}.

\section{Electroweak penguins.}\label{sec:ewpenguins}

There is another instance where our understanding of the
$\Pi_{LR}$ function will help us.  This is in the
$\mathcal{O}(p^0)$ calculation of the electroweak penguin $Q_8$,
which is defined as
\begin{equation}\label{Q8}
  Q_8= - 12 e\ (\lambda^{(32)}_L)_{ij}\ (Q_R)_{kl}\ \overline{q'^i_L} {q^k_R}(x)\
  \overline{q^l_R} {q'^j_L}(x)\ ,
\end{equation}
where $Q_R$ is the same as in the previous section and
$(\lambda^{(32)}_L)_{ij}=\delta_{i3}\delta_{j2}$. This operator
is the same as $D_{LR}$ in Eq. (\ref{D}) except that the
$\lambda_{L}$ matrix replaces now $Q_L$. Therefore following our
previous discussion one can see that $Q_8$ bosonizes at
$\mathcal{O}(p^0)$ as\cite{KPdeR01}
\begin{equation}\label{bosonQ8}
Q_8= - 12\ <0|\overline{s}_L s_R(0)\ \overline{d}_R d_L(0) |0> \
\mathrm{tr}\left(U\lambda_L^{(23)}U^{\dag}Q_R\right)^{\dag}\ ,
\end{equation}
where, unlike in the previous section, we have not approximated
the above four-quark condensate by its factorized expression.
This we do as a cautionary measure, just in case subleading terms
in the large-$N_c$ expansion turn out to be of numerical
importance in the Zweig-rule violating scalar amplitudes, as
sometimes is claimed in the literature\cite{Moussallam}.

As it turns out, it is precisely this four quark condensate in
Eq. (\ref{bosonQ8}) which governs the $1/Q^6$ fall-off of the OPE
of $\Pi_{LR}(Q^2)$ in the chiral $SU(3)$ limit. Specifically  one
finds that
\begin{equation}\label{falloff}
  -Q^6 \Pi_{LR}(Q^2) = 16 \pi \alpha_s
  \left(1 + \xi \frac{\alpha_s}{\pi}\right)
  <0|\overline{s}_L s_R\ \overline{d}_R d_L |0>+\cdots  +
  \mathcal{O}(\frac{1}{Q^2})\ ,
\end{equation}
where the $\cdots$ stand for smaller contributions\cite{KPdeR01}.
The term $\xi$ was not known at the time we wrote our
paper\cite{KPdeR01}, but now it is\cite{Cirigliano,Bijnens}:\
$\xi=25/8$ in NDR, and $\xi=21/8$ in HV. Using the MHA with just
the $\rho$ and the $a_1$ as we did in the case of the
$\pi^+-\pi^0$ mass difference, this large-$Q^2$ fall-off is given
by
\begin{equation}\label{falloffMHA}
  -Q^6 \Pi_{LR}(Q^2)\quad
  {\longrightarrow}^{^{^{\!\!\!\!\!\!\!\!\!\!\!\!\!\!\!\!Q^2 \rightarrow
  \infty}}} \quad f_A^2 M_A^2-f_V^2 M_V^2 \ =\  f_{\pi}^2 M_V^2
  M_A^2\ .
\end{equation}
Choosing $\alpha_s(2\ GeV)\simeq 0.33\pm 0.04$ one obtains a
value for the four quark condensate in Eq. (\ref{falloff}) and
with it and Eq. (\ref{bosonQ8}) one can compute, e.g., the matrix
elements $M_{8}\equiv<(\pi\pi)_{I=2}|Q_{8}|K^0>(2\ GeV)$. One can
proceed analogously with the other EW penguin $Q_7$, which is
akin to $Q_{LR}$ in Eq. (\ref{Zeff}), and the corresponding matrix
element $M_7$\cite{KPdeR01}. An updated summary of results for
these matrix elements is given in Table 1, where I have limited
myself to those theoretical approaches which are capable of
matching to the Wilson coefficients at short
distances\footnote{The Trieste group has also obtained values for
$M_{8,7}$\cite{Trieste}, although without this matching.}.
\begin{table}[t]
\caption{Summary of results for
$M_{7,8}\equiv<(\pi\pi)_{I=2}|Q_{7,8}|K^0>(2\ GeV)$, in units of
$GeV^3$.}
\begin{center}
\footnotesize
\begin{tabular}{|c|c|c|c|c|}
  \hline
Refs. & $M_7$(NDR) & $M_7$(HV) & $M_8$(NDR) & $M_8$(HV)
 \\\hline
 Knecht et al.\cite{KPdeR01} & $0.11\pm 0.03$ & $0.67\pm0.20$ & $2.34\pm0.73$ & $2.52\pm0.79$ \\
 Narison\cite{Narison} & $0.35\pm 0.10$ &  & $2.7\pm 0.6$ &  \\
Cirigliano et al.\cite{Cirigliano} & $0.16\pm0.10$ & $0.49\pm 0.07$ & $2.22\pm0.67$ & $2.46\pm 0.70$ \\
 Bijnens et al.\cite{Bijnens} & $0.24\pm 0.03$ & $0.37\pm 0.08$ & $1.2\pm 0.9$ & $1.3\pm 0.9$ \\
 Battacharya et al.\cite{Batta} & $0.32\pm 0.06$ &  & $1.2\pm 0.2$ &  \\
 Donini et al.\cite{Donini} & $0.11\pm 0.04$ & $0.18\pm 0.06$ & $0.51\pm 0.10$ & $0.62\pm 0.12$
  \\ \hline
\end{tabular}
\end{center}
\end{table}

 To conclude I would like to emphasize that the MHA
has also been successfully applied to other processes such as
$\pi^0\rightarrow e^+e^-$ and $\eta \rightarrow
\mu^+\mu^-$\cite{eta} and has been very instrumental in recent
reanalyses of $g_{\mu}-2$\cite{g}. Further calculations are in
progress.

\section*{Acknowledgements}
I would like to thank the organizers for their kind invitation to
this Conference and M. Golterman, T. Hambye, M. Knecht, A.
Nyffeler, M. Perrottet and E. de Rafael for a very enjoyable
collaboration and interesting discussions. Work supported by
CICYT-AEN99-0766, 2001 SGC 00188 and by TMR, EC-Contract No.
ERBFMRX-CT980169 (Eurodaphne).

\end{document}